\documentclass[twocolumn,showpacs,preprintnumbers,amsmath,amssymb,floatfix]{revtex4}
\pdfoutput=1

\usepackage{graphicx}
\usepackage{dcolumn}
\usepackage{bm}
\usepackage{amsmath}
\usepackage{color}

\begin{document}

\newcommand{\ins}[1]{\textcolor{blue}{#1}}
\newcommand{\del}[1]{\textcolor{red}{#1}}

\newcommand{\Rphia}[1] { \varphi_a \! \left( #1 \right) }
\newcommand{\Rphim}[1] { \varphi_m \! \left( #1 \right) }
\newcommand{\Rphiastar}[1] { \varphi_a^* \! \left( #1 \right) }
\newcommand{\Rphimstar}[1] { \varphi_m^* \! \left( #1 \right) }
\newcommand{\Rphiasq}[1] { \varphi_a^2 \! \left( #1 \right) }
\newcommand{\Rphimsq}[1] { \varphi_m^2 \! \left( #1 \right) }
\newcommand{\Rphiastarsq}[1] { \varphi_a^{*2} \! \left( #1 \right) }
\newcommand{\Rphimstarsq}[1] { \varphi_m^{*2} \! \left( #1 \right) }
\newcommand{\RGN}[2] { \mathcal{G}_N \! \left( #1, #2 \right) }
\newcommand{\RGA}[2] { \mathcal{G}_A \! \left( #1, #2 \right) }
\newcommand{\RGNI}[1] { \mathcal{G}_N^I \! \left( #1 \right) }
\newcommand{\RGNAI}[1] { \mathcal{G}_{N/A}^I \! \left( #1 \right) }
\newcommand{\RGAI}[1] { \mathcal{G}_A^I \! \left( #1 \right) }
\newcommand{\RGNstar}[2] { \mathcal{G}_N^* \! \left( #1, #2 \right) }
\newcommand{\RGAstar}[2] { \mathcal{G}_A^* \! \left( #1, #2 \right) }
\newcommand{\RGNIstar}[1] { \mathcal{G}_N^{* \, I} \! \left( #1 \right) }
\newcommand{\RGAIstar}[1] { \mathcal{G}_A^{* \, I} \! \left( #1 \right) }
\newcommand{\RGNlm}[2] { \mathcal{G}_N^{#1,#2} \! \left( R, k \right) }
\newcommand{\RGAlm}[2] { \mathcal{G}_A^{#1,#2} \! \left( R, k \right) }
\newcommand{\RGNn}[1] { \mathcal{G}_N^{#1} \! \left( R_{\rho}, R_z, k_{\rho}, k_z \right) }
\newcommand{\RGNAn}[1] { \mathcal{G}_{N/A}^{#1} \! \left( R_{\rho}, R_z, k_{\rho}, k_z \right) }
\newcommand{\RGAn}[1] { \mathcal{G}_A^{#1} \! \left( R_{\rho}, R_z, k_{\rho}, k_z \right) }
\newcommand{\Rcyls}[0] { R_{\rho}, R_z }
\newcommand{\kcyls}[0] { k_{\rho}, k_z, \phi }
\newcommand{\Rhalf}[0] { R_{\rho}^{1/2} }
\newcommand{\wrt}[1]{ \mathrm{d}#1 \; }
\newcommand{\wrttwo}[1]{ \mathrm{d}^2 #1 \; }
\newcommand{\wrtthree}[1]{ \mathrm{d}^3 #1 \; }
\newcommand{\Vabar}[1] { \bar{V}_a \! \left( #1 \right) }
\newcommand{\Vmbar}[1] { \bar{V}_m \! \left( #1 \right) }
\newcommand{\parpar}[1] { \frac{\partial}{\partial #1} }
\newcommand{\modsq}[1] { \left| #1 \right|^2 }
\newcommand{\parpartwo}[1] { \frac{\partial^2}{\partial #1^2} }

\preprint{}

\title{{H}artree-{F}ock-{B}ogoliubov Model and Simulation of Attractive and Repulsive {B}ose-{E}instein Condensates}

\author{V.\ D.\ Snyder$^1$}
\author{S.\ J.\ J.\ M.\ F. Kokkelmans$^2$}
\author{L.\ D.\ Carr$^{1,3}$}%
\affiliation{%
$^1$Department of Physics, Colorado School of Mines, Golden, CO 80401, USA \\
$^2$Eindhoven University of Technology, P.O. Box 513, 5600 MB Eindhoven, The
Netherlands \\
$^3$Ruprecht-Karls-Universit\"at Heidelberg, Physikalisches Institut, Philosophenweg 12, D-69120 Heidelberg, Germany
}%


\date{\today}

\begin{abstract}
We describe a model of dynamic Bose-Einstein condensates near a Feshbach resonance that is computationally feasible under assumptions of spherical or cylindrical symmetry.  Simulations in spherical symmetry approximate the experimentally measured time to collapse of an unstably attractive condensate only when the molecular binding energy in the model is correct, demonstrating that the quantum fluctuations and atom-molecule pairing included in the model are the dominant mechanisms during collapse.  Simulations of condensates with repulsive interactions find some quantitative disagreement, suggesting that pairing and quantum fluctuations are not the only significant factors for condensate loss or burst formation.  Inclusion of three-body recombination was found to be inconsequential in all of our simulations, though we do not consider recent experiments \cite{AltinEtAl2011} conducted at higher densities.
\end{abstract}

\pacs{03.75.Kk, 67.60.Bc, 67.10.Jn}
\maketitle

\section{Introduction\label{IntroSec}}

Bose-Einstein condensates (BECs) which are suddenly subjected to strong
attractive interatomic interactions, can undergo exotic collapse that resembles
supernovas~\cite{Donley2001}.  Experiments \cite{Claussen2002,Donley2002}
performed in the seemingly opposite vein, using strong repulsive interactions,
exhibit some of the same features of the collapse; namely, an energetic burst of
atoms, a remnant condensate, and a significant portion of the atoms escaping
detection.  These particular experimental observations of collapsing BECs have eluded satisfying quantitative explanation for years, with the time to collapse being particularly difficult to reproduce in simulations.

The regime of strong interatomic interactions is reached by using a Feshbach resonance, where an external magnetic field allows for tuning the sign and strength of the scattering length.
We present a theory of BECs near a Feshbach resonance, including lowest-order fluctuations.  Since a collapse is a highly local effect, it is crucial that we allow for strong inhomogeneities during the simulation.  This model is defined over time and six spatial variables. Symmetry assumptions and the restriction of our knowledge to only certain off-diagonal correlations reduces the model to four or five spatial degrees of freedom.  Assuming spherical symmetry, our simulations predict a collapse time of about 2 milliseconds for one of the condensates described in Ref.\ \cite{Donley2001}, which agrees well with the experiment.  Simulations of experiments \cite{Claussen2002} on condensates with repulsive interatomic interactions consistently overestimate the number of atoms remaining after a brief period near the Feshbach resonance.  In these repulsive simulations, we observe that bound pairs attain high velocities (above 8 millimeters per second) immediately before dissociating.  In all simulations, inclusion of an empirically-based model of three-body recombination had no significant effect.

We address the experimentally measured time to collapse, which has thus far not
been accurately reproduced for the particular experiments we model.  We also explore elements of the dynamics that have received little experimental or theoretical attention, such as the kinetics of bound pairs during collapse, and attempt to reproduce the results of experiments with a single pulse near the resonance.

Section \ref{OverviewSec} describes the experiments and summarizes past simulations.  Section \ref{ModelSec} derives the Hartree-Fock-Bogoliubov model and performs the simplifications needed to make the resulting equations practical for simulation.  We present the results of our simulations in Section \ref{ResultsSec} and draw conclusions in Section \ref{ConcSec}.

\section{\label{OverviewSec}Overview of Experiment and Theory of Collapse and Related Dynamics}

\subsection{\label{ExperimentSubsec}Experiment}

By exploiting a Feshbach resonance \cite{ChinEtAl2010}, the interactions between condensed atoms can be tuned from repulsive to attractive values over only a few microseconds \cite{Donley2002}.  In an often-examined set of experiments \cite{Donley2001, Roberts2001, ClaussenTh} conducted at JILA, condensates of about 15,000 $^{85}$Rb atoms were formed at a temperature of 3 nanokelvin with slightly repulsive interactions.  The repulsion was balanced by a magnetic trap that was well approximated by an axisymmetric harmonic potential, so that an initial condensate was stable and neither expanding nor collapsing.  The interactions were then suddenly tuned to be attractive.  A condensate would appear stable for a short time (the \emph{collapse time} $t_\text{collapse}$) after this transition, then suddenly lose atoms at an exponential rate.  During collapse, a burst of energetic atoms was emitted from the condensate.  Between experiments, the number of atoms in the bursts varied by as much as a factor of two, even for identical sets of controlled and observed experimental parameters.  If the atom loss was interrupted by changing the strength of the interactions a second time, now to a slightly repulsive value, jets of atoms having a lower energy than the bursts were emitted, almost entirely in the radial direction.  A stable, excited, and highly anisotropic condensate remained after atom loss ceased.

A significant number of atoms lost from the condensate went undetected; for example, about 8500 atoms out of an initial condensate of 15,000 were missing after a collapse \cite{ClaussenTh}.  Atoms with energies greater than about 20 $\mathrm{\mu}$K, atoms in states that were not influenced by the trapping potential, and pairs of atoms bound to each other because of the Feshbach resonance (henceforth referred to as \emph{molecules}) could not be detected.

One set of related experiments \cite{Claussen2002, ClaussenTh} started from a stable, noninteracting or weakly repulsive BEC, which was subjected to a rapid magnetic field pulse.  The field was linearly ramped from its initial value to a value near the Feshbach resonance in tens to hundreds of microseconds, held at a constant value for one to hundreds of microseconds, called the \emph{hold time}, and then quickly and linearly ramped back to the initial value.  Likewise, the scattering length was ramped from zero or a small positive value, to a very large positive value, and finally back to its initial small value.

Following a pulse, it was observed that the number of atoms remaining in the condensate increased for longer ramp times, indicating that the dominant loss mechanism was not the usual density-dependent loss responsible for the rethermalization of a stable condensate.  Varying the initial density of the BEC did not appreciably alter the rate of loss, further suggesting that the loss was not density-dependent.  As expected, pulses which came closer to the resonance resulted in more loss from the condensate.  When the scattering length was held at a large positive value during the hold time, small, damped oscillations in atom number were apparent when the hold time was varied.

A burst of atoms similar to that in the collapse experiments appeared in these experiments with repulsive interactions.  For small positive values of the scattering length, no burst atoms were observed.  Varying the number of atoms in the surrounding thermal cloud did not appreciably affect the bursts, indicating that interactions with noncondensed atoms are not responsible. The burst atoms remained in the same spin state as the condensed atoms, suggesting that spin-flip interactions were not involved.

These single-pulse experiments inspired experiments \cite{Donley2002, ClaussenTh} with two magnetic field pulses, separated by a \emph{free precession} time, during which the magnetic field was held constant and below the initial value.  As with the other scenarios, an energetic burst of atoms emanated from the condensate.  Again, between 8 and 50 percent of the atoms escaped detection.

\subsection{\label{TheorySubsec}Theory}

BEC collapse has been theoretically studied for several years.  Most recently,
Altin \emph{et al}.\ \cite{AltinEtAl2011} performed new collapse measurements
in an optical trap, with a $^{85}$Rb condensate that was much denser than those
of the JILA experiments.  In this regime, they found a mean field description in
combination with three-body losses gave a good description of their measured
collapse time and atom loss curve.  In our paper, we focus on the much less
understood JILA experiments, where three-body losses are shown to play an
inconsequential role.

Kagan \emph{et al}.\ \cite{Kagan1997} predicted that collapse occurs on a time scale $t_{\text{collapse}} \sim \omega^{-1}$, where $\omega$ is the trap frequency.  The observations of \cite{Donley2001} have shown this prediction to be incorrect.  Kagan and coworkers also supposed \cite{Kagan1998} that during a collapse, the condensate's density increases until density-dependent losses due to three-body recombination take over, eventually causing expansion of the condensate.  The cycle then repeats, as the trap pushes the remaining condensate back towards the trap center.  The GPE simulations of Saito and Ueda and Bao \cite{Saito2001, Saito2001a, Saito2002, Bao2004} clearly show such behavior, leading to significant atom loss and the prevention of the appearance of a singularity during collapse.

These and other \cite{Adhikari2002, Adhikari2004, Adhikari2005, Santos2002} simulations qualitatively reproduce the collapse process, the delay before atom loss begins, the condensate number decay constant $\tau_{\text{decay}}$, bursts, and jets, but have achieved no solid quantitative agreement with observation.  Minor differences in these authors' results, as well as the lack of quantitative agreement with experiment, may be due to their different choices of density-dependent loss rates.  These losses have been shown \cite{Roberts2000} to have a complicated dependence on magnetic field, especially near a Feshbach resonance, making them difficult to precisely characterize.

Recognizing the deficiency in atom loss models, Bao and coworkers \cite{Bao2004} performed a GPE simulation with a loss rate chosen so that their simulations correctly reproduced the experimental values of $t_{\text{collapse}}$ and condensate remnant number.  The atom number decay constant $\tau_{\text{decay}}$ is reasonably well reproduced, but the simulated burst energies are much lower than what is experimentally observed.  This failure suggests that a Gross-Pitaevskii model with simple density-dependent loss does not sufficiently describe the collapse.  Savage \emph{et al}.\ \cite{Savage2003}, surveying the literature and performing their own simulations with several different loss rate coefficients, arrive at the same conclusion, noting that theoretical values of $t_{\text{collapse}}$ are consistently larger than the experimental values.  The authors mention that this is surprising, since the period before collapse begins should be the domain where the GPE applies.

Duine and Stoof \cite{Duine2001} propose that two condensed atoms can collide, scattering one atom out of the condensate.  They use a Gaussian variational technique to investigate this ``quantum evaporation" as a possible player in the collapse, especially concerning remnant number and burst formation.  Their simulations show a considerable loss from the condensate, but, disagreeing with observation, this loss begins immediately after the interatomic interactions become attractive.

Mackie and coworkers \cite{Mackie2002} suggest that pairs formed by the Feshbach resonance may dissociate into noncondensed atoms during the collapse, and the simulations of Milstein \emph{et al}.\ \cite{Milstein2003}, which neglect three-body losses but include quantum fluctuations and pair formation via the Feshbach resonance, show an energetic burst of noncondensed atoms, though using parameters not taken from experiments.

Calzetta \emph{et al}.\ \cite{Calzetta2003} downplay the importance of such a molecular component for the values of the scattering length $a_{\text{collapse}}$ in the collapse experiments, which are far from resonance.  Like Yurovsky \cite{Yurovsky2002}, they attribute loss from the condensate to the growth of noncondensed modes.  Calzetta \emph{et al}.\ suggest that a theory accounting for fluctuations would have instabilities growing out of those fluctuations, which may account for earlier collapse times, or that a loss of coherence between atoms is largely responsible for atom loss \cite{Calzetta2008}.

W\"{u}ster and coworkers \cite{Wuster2005} use the same theory of fluctuations as Milstein \emph{et al}.\, but also regard the molecular component as unimportant to collapse.  Their simulations still find a $t_{\text{collapse}}$ exceeding the observed value.  Using an alternate, truncated Wigner formulation and including initial and dynamical noise, a background thermal component, and cylindrical geometry with experimental parameters, W\"{u}ster \emph{et al}.\ \cite{Wuster2007} still overestimate the experimentally measured $t_{\text{collapse}}$ by about 40 percent.

Haldar \emph{et al}.\ \cite{HaldarEtAl2010} summarize a correlated potential harmonics expansion method that accounts for two-body correlations and models interatomic interactions with the van der Waals potential.  They show that anharmonicity and a finite potential barrier at the ends of the optical trap have a nontrivial effect on the stability of attractive condensates.  The same method is used to demonstrate a variation of energies with effective scattering length where mean field theory predicts none \cite{DebnathEtAl2011}, and predicts the critical number of condensed atoms at which a condensate collapses much more accurately than does mean field theory \cite{DasEtAl2009}, highlighting the importance beyond-mean-field effects in collapsing condensates.

All these models have at least some qualitative agreement with observation, and some provide insight into other aspects of the collapse experiments \cite{Saito2002, Bao2004, Adhikari2004}.  Saito and Ueda \cite{Saito2001} suggest the bursts are atoms originally near the center of the collapse that acquire kinetic energy when three-body losses suddenly remove a large number of atoms from the center of the collapse.  In these simulations and others \cite{Saito2002, Adhikari2002, Adhikari2004, Adhikari2005, Santos2002, Bao2004}, the burst atoms are distinguished from the condensate purely by their location.  In the simulations of Milstein \emph{et al}.\ and W\"{u}ster \emph{et al}.\, the burst is assumed to be a distinct noncondensed field, which can occupy the same space as the condensate.

The magnetic field pulse experiments have stimulated fewer simulations than the collapse experiments.  Duine and Stoof \cite{Duine2003} use coupled mean fields allowing for molecule formation, quantum evaporation, and three-body losses in modeling the one-pulse experiments.  These simulations had only general qualitative agreement with the experiments, but with the observation that the inclusion of three-body losses supressed oscillations in numbers of atoms and molecules, despite the belief that these density-dependent losses should be unimportant under the experimental circumstances \cite{ClaussenTh}.  Mackie and coworkers \cite{Mackie2002} use a coupled mean field model that allows for dissociation of molecules into noncondensed atom pairs, but find only about five percent loss to the noncondensed component, with very few molecules being retained.  The authors observe a larger loss in simulations of the two-pulse experiments, but the oscillation envelopes have a behavior markedly different from the slight damping observed in the experiments.  Kokkelmans and Holland \cite{Kokkelmans2002a} use the same model as the Milstein \emph{et al}.\ collapse simulation \cite{Milstein2003}, but use a Gaussian average over a homogeneous gas to simulate the behavior of a trapped gas.  These simulations agreed fairly well with the two-pulse experiments, showing that the majority of atoms lost from the condensate go into noncondensed modes, and the missing atoms are identified as molecules.    K\"{o}hler and coworkers \cite{Kohler2003} model the two-pulse experiments with a theory that includes molecule formation and quantum fluctuations, and find good qualitative agreement with the experiments.  They also find that the presence of the trap moves the means of the condensate and burst numbers' oscillations closer together in a way not captured by a Gaussian average of a homogeneous gas.  They attribute this difference to the presence of a length scale not found in the homogeneous gas simulations.

Many of the questions posed by the experiments have eluded satisfying explanation.  In the case of collapse, the experimentally measured $t_\text{collapse}$ has been particularly difficult to simulate.  Unsettled points of contention include the mechanisms by which the jets and bursts operate, and the importance of three-body losses to the collapse.  The counter-intuitive behavior of the condensate after the collapse has thus far received relatively little attention \cite{Parker2008}, as have the simulations involving a single pulse of repulsive interactions.  There has also been little exploration of the various models' implications for measurable quantities and phenomena that have not yet been vigorously pursued in experiments.

\section{\label{ModelSec}Hartree-Fock-Bogoliubov Model}

To treat both collapse experiments and single and double pulse purely repulsive
experiments we will work with a Hartree-Fock-Bogoliubov (HFB) model which
explicity takes into account the two main channels involved with the Feshbach
resonance.  Previous versions of this model exist \cite{Holland2001,
Milstein2003}, including operator equations and factorized expectation values of
the HFB equations \cite{Kokkelmans2002, Wuster2005}, but we present adaptations
to spherical and cylindrical geometry in a complete and rigorous derivation, as not found elsewhere to our knowledge.

We begin with a definition of the atomic field operator as
\begin{equation}
\hat{\psi}_a \! \left( \mathbf{x} \right) = \phi_a \! \left( \mathbf{x} \right) + \hat{\chi} \! \left( \mathbf{x} \right),
\label{Decomp}
\end{equation}
where $\phi_a \! \left( \mathbf{x} \right) \equiv \langle \hat{\psi}_a \! \left( \mathbf{x} \right) \rangle$, and all operators are taken to be in the Heisenberg picture, though for brevity we do not explicitly write the time dependence.  We call a pair of atoms in a quasibound state due to the Feshbach resonance a \emph{molecule}.  We loosely assign the term ``molecule" to the closed channel for the purpose of this discussion.  The actual molecule is a superposition of open and closed channels.  The field operator for molecules is represented by $\hat{\psi}_m \! \left( \mathbf{x} \right)$, with expectation value $\phi_m \! \left( \mathbf{x} \right) \equiv \langle \hat{\psi}_m \! \left( \mathbf{x} \right) \rangle$.  To keep the problem tractable, we assume no fluctuations around the molecular condensate.  The normal and anomalous fluctuations are defined by
\begin{align}
G_N \! \left( \mathbf{x}, \mathbf{x'} \right) & \equiv \langle \hat{\chi}^\dagger \! \left( \mathbf{x'} \right) \hat{\chi} \! \left( \mathbf{x} \right) \rangle \nonumber \\
& = \langle \hat{\psi}_a^\dagger \! \left( \mathbf{x'} \right) \hat{\psi}_a \! \left( \mathbf{x} \right) \rangle - \langle \hat{\psi}_a^\dagger \! \left( \mathbf{x'} \right) \rangle \langle \hat{\psi}_a \! \left( \mathbf{x} \right) \rangle \nonumber \\
& = \langle \hat{\psi}_a^\dagger \! \left( \mathbf{x'} \right) \hat{\psi}_a \! \left( \mathbf{x} \right) \rangle - \phi_a^* \! \left( \mathbf{x'} \right) \phi_a \! \left( \mathbf{x} \right)
\end{align}
and
\begin{align}
G_A \! \left( \mathbf{x}, \mathbf{x'} \right) & \equiv \langle \hat{\chi} \! \left( \mathbf{x'} \right) \hat{\chi} \! \left( \mathbf{x} \right) \rangle \nonumber \\
& = \langle \hat{\psi}_a \! \left( \mathbf{x'} \right) \hat{\psi}_a \! \left( \mathbf{x} \right) \rangle - \langle \hat{\psi}_a \! \left( \mathbf{x'} \right) \rangle \langle \hat{\psi}_a \! \left( \mathbf{x} \right) \rangle \nonumber \\
& = \langle \hat{\psi}_a \! \left( \mathbf{x'} \right) \hat{\psi}_a \! \left( \mathbf{x} \right) \rangle - \phi_a \! \left( \mathbf{x'} \right) \phi_a \! \left( \mathbf{x} \right)
\end{align}
respectively.  The diagonal elements ($\mathbf{x} = \mathbf{x'}$) of the normal fluctuations give a number density of noncondensed atoms, while the diagonal elements of the anomalous fluctuations give the variance in the mean $\phi_a \! \left( \mathbf{x} \right)$.  Off diagonal elements of the normal and anomalous fluctuations are equal-time correlation functions.

We obtain equations of motion for the atomic and molecular mean fields and the normal and anomalous fluctuations by finding Heisenberg equations of motion for the atomic and molecular field operators and for the products $\hat{\chi}^\dagger \! \left( \mathbf{x'} \right) \hat{\chi} \! \left( \mathbf{x} \right)$ and $\hat{\chi} \! \left( \mathbf{x'} \right) \hat{\chi} \! \left( \mathbf{x} \right)$. Taking the expectation value on both sides of the resulting equations results in averages of products of atomic and molecular field operators; by assuming that the atomic and molecular field operators act in orthogonal subspaces of the full Hilbert space, these expectation values factorize as, for example,
\begin{equation}
\left\langle \hat{\psi}_a^\dagger \! \left( \mathbf{x} \right) \hat{\psi}_m \! \left( \mathbf{x'} \right) \right\rangle = \phi_a^* \! \left( \mathbf{x} \right) \phi_m \! \left( \mathbf{x'} \right).
\label{AMFactEx}
\end{equation}
Products of three or four atomic field operators, which do appear in the equations of motion, require special care to be expressed in terms only of the atomic mean field and normal and anomalous fluctuations.

We may apply a manifestation of Wick's theorem \cite{BlaizotRipka} if the state of the system is an eigenstate of every Bogoliubov atomic quasiparticle annihilation operator (a linear superposition of the atomic momentum-space creation and annihilation operators).  This quasiparticle coherent state, which is not generally a coherent state of the field operator $\hat{\psi}_a \! \left( \mathbf{x} \right)$, is a squeezed state \cite{LeBellac}, and thus saturates the number-phase Heisenberg uncertainty relation.  Then
\begin{align}
\left\langle \hat{\psi}_a^\dagger \! \left( \mathbf{x} \right) \hat{\psi}_a \! \left( \mathbf{x} \right) \hat{\psi}_a \! \left( \mathbf{x} \right) \right\rangle = & \left| \phi_a \! \left( \mathbf{x} \right) \right|^2 \phi_a \! \left( \mathbf{x} \right) + \phi_a^* \! \left( \mathbf{x} \right) G_A \! \left( \mathbf{x}, \mathbf{x} \right) \nonumber \\
& + 2 \phi_a \! \left( \mathbf{x} \right) G_N \! \left( \mathbf{x}, \mathbf{x} \right),
\label{WickEx}
\end{align}
for example, is exact.

We use the Hamiltonian
\begin{align}
\hat{H} = & \int \mathrm{d}^3 x'' \, \hat{\psi}_a^\dagger \! \left( \mathbf{x''} \right) \left[ -\frac{\hbar^2}{2m} \nabla^2 + V \! \left( \mathbf{x''} \right) \right] \hat{\psi}_a \! \left( \mathbf{x''} \right) \nonumber \\
& + \int \mathrm{d}^3 x'' \, \hat{\psi}_m^\dagger \! \left( \mathbf{x''} \right) \left[ -\frac{\hbar^2}{4m} \nabla^2 + 2 V \! \left( \mathbf{x''} \right) + \nu \right] \hat{\psi}_m^\dagger \! \left( \mathbf{x''} \right) \nonumber \\
& + \frac{U}{2} \int \mathrm{d}^3 x'' \, \hat{\psi}_a^\dagger \! \left( \mathbf{x''} \right) \hat{\psi}_a^\dagger \! \left( \mathbf{x''} \right) \hat{\psi}_a \! \left( \mathbf{x''} \right) \hat{\psi}_a \! \left( \mathbf{x''} \right) \nonumber \\
& + \frac{g}{2} \int \mathrm{d}^3 x'' \left[ \hat{\psi}_m^\dagger \! \left( \mathbf{x''} \right) \hat{\psi}_a \left( \mathbf{x''} \right) \hat{\psi}_a \left( \mathbf{x''} \right) + \text{h.c.} \right],
\label{Ham}
\end{align}
which is often referred to as a two-channel Hamiltonian, where $m$ is the mass of an atom, $V \! \left( \mathbf{x''} \right)$ is the external potential felt by a single atom, $\nu$ is the detuning of the Feshbach resonance, $U$ relates to the strength of the non-resonant atom-atom interaction, and $g$ relates to the strength of the atom-molecule coupling which gives rise to the Feshbach resonance. These interaction parameters are based on the assumption of contact interactions between the particles. Such interactions, however,  give rise to an ultraviolet divergence in momentum space, which must be treated properly by renormalization. This is done by the introduction of a momentum cutoff $K$, while making sure that at the same time the correct underlying two-body resonance physics is maintained. Therefore it is necessary to consider the contact interactions as the zero range limits of the actual nonlocal interactomic potentials. The properties of the contact potentials can then be chosen such that the two-body physics around a Feshbach resonance is correctly described~\cite{Kokkelmans2002,Kokkelmans2002a}. This renormalization procedure amounts to a $K$-dependent relationship between the interaction parameters in the Hamiltonian and the physical interaction parameters given by
\begin{align}
U & = \Gamma U_0, \nonumber \\
g & = \Gamma g_0, \nonumber \\
\nu & = \nu_0 + \frac{1}{2} \alpha g g_0,
\end{align}
where
\begin{equation}
\Gamma \equiv \frac{1}{1 - \alpha U_0} \text{ and } \alpha \equiv \frac{m K}{2 \pi^2 \hbar^2}.
\end{equation}
The parameters with a subscript $0$ are the unrenormalized physical interaction parameters and are defined by
\begin{align}
U_0 & \equiv \frac{4 \pi \hbar^2 a_\text{bg}}{m}, \nonumber \\
g_0 & \equiv \sqrt{g_c U_0 \, \Delta B \, \Delta \mu^\text{mag}}, \nonumber \\
\nu_0 & \equiv \left( B - B_\text{res} \right) \Delta \mu^\text{mag},
\label{unrenormedParams}
\end{align}
where $a_{\text{bg}}$ is the background scattering length; $\Delta B$ is the width of the Feshbach resonance, defined as the distance from the resonance position to the point where the effective scattering length is zero; $\Delta \mu^{\text{mag}}$ is the difference in magnetic moments between an uncoupled bound and unbound pair of atoms; $B$ is the external magnetic field; $B_{\text{res}}$ is the position of the resonance, defined as the value of the magnetic field for which the effective scattering length diverges; and the correction factor $g_c$ may be set to $1.816$ to match the binding energy of the contact potential model as closely as possible to the field-dependent binding energy of the weakly-bound rubidium Feshbach state~\cite{Kokkelmans2002a}, or to $2$ to match the mean field energy.  The values of those parameters which are fixed are summarized in Table \ref{ParameterTable}.  The cutoff $K$ is set to $6 \times 10^{8}$ inverse meters, the largest wavenumber in our simulations.  One may combine the unrenormalized parameters \eqref{unrenormedParams} and $a_\text{eff} = a_\text{bg} \left[ 1 - \Delta B / \left( B - B_\text{res} \right) \right]$ to write
\begin{equation}
a_\text{eff} = a_\text{bg} - \frac{m}{4 \pi \hbar^2} \frac{g_0^2}{g_c \nu_0}.
\label{aEffFromNu0}
\end{equation}

\begin{table}
\begin{center}
\begin{tabular}{|c|c|}
\hline
Parameter & Value \\
\hline
$m$ & $84.911794 m_\text{proton}$ \\
\hline
$a_\text{bg}$ & $-450.0 a_0$ \\
\hline
$\Delta B$ & $10.95$ G \\
\hline
$\Delta \mu^\text{mag}$ & $-2.2259 \mu_B$ \\
\hline
$B_\text{res}$ & $154.9$ G \\
\hline
\end{tabular}
\caption{Values of the fixed renormalization parameters that we used in our code, where $m_\text{proton}$ is the mass of a proton, $a_0$ is the Bohr radius, and $\mu_B$ is the Bohr magneton.\label{ParameterTable}}
\end{center}
\end{table}

Consistent with the two-body approximation, the Hamiltonian (\ref{Ham}) neglects all interactions between atoms and molecules that do not involve molecule formation or dissociation, and assumes molecules do not interact with each other.

When we transform to the center of mass $\mathbf{R} \equiv \left( \mathbf{x} + \mathbf{x'} \right)/2$ and relative $\mathbf{r} \equiv \mathbf{x} - \mathbf{x'}$ coordinates, we can write
\begin{align}
\bar{\phi}_a \! \left( \mathbf{R} \right) & \equiv \phi_a \! \left( \mathbf{x} \right) = \phi_a \! \left( \mathbf{x'} \right), \nonumber \\
\bar{\phi}_m \! \left( \mathbf{R} \right) & \equiv \phi_m \! \left( \mathbf{x} \right) = \phi_m \! \left( \mathbf{x'} \right), \nonumber \\
\bar{G}_N \! \left( \mathbf{R}, \mathbf{r} \right) & \equiv G_N \left( \mathbf{x}, \mathbf{x'} \right), \nonumber \\
\bar{G}_A \! \left( \mathbf{R}, \mathbf{r} \right) & \equiv G_A \left( \mathbf{x}, \mathbf{x'} \right), \nonumber \\
\bar{V} \! \left( \mathbf{R} \right) & \equiv V \! \left( \mathbf{x} \right) = V \! \left( \mathbf{x'} \right). \label{CoMNotn}
\end{align}
These statements will be valid if, in the case of spherical geometry, the external potential and the initial conditions on all single-particle fields (diagonal elements of $G_N$ and $G_A$ included) are rotationally invariant, and we only consider $\mathbf{x}$ and $\mathbf{x'}$ such that
\begin{equation}
\left| \mathbf{x} \right| = \left| \mathbf{x'} \right|.
\label{SphRestr}
\end{equation}
For cylindrical geometry, we assume the initial conditions on single-particle fields and the external potential are invariant with respect to rotation about a vertical axis (let it be the $z$-axis) and invariant with respect to reflections over a plane normal to that axis (the $x\text{-}y$ plane); we must also restrict $\mathbf{x}$ and $\mathbf{x'}$ such that
\begin{align}
\left| \mathbf{x}_\rho \right| & = \left| \mathbf{x}_\rho' \right|, \nonumber \\
\left| \mathbf{x}_z \right| & = \left| \mathbf{x}_z' \right|, \label{CylRestr}
\end{align}
where $\mathbf{x}_\rho$ is the component of $\mathbf{x}$ lying in the $x\text{-}y$ plane and $\mathbf{x}_z$ is the component along the $z$-axis.
Note that the restrictions (\ref{SphRestr}) and (\ref{CylRestr}) impose no additional approximations beyond those that have already been made.  They merely provide convenient simplifications in the equations of motion.

The notation in Eq.\ (\ref{CoMNotn}) is general enough to handle the spherical and cylindrical cases, though only the magnitude of $\mathbf{R}$ is important to single-particle fields in the former, and only the magnitudes of $\mathbf{R}_z$ and $\mathbf{R}_\rho$ are important to single-particle fields in the latter.  For consistency, we make the same assumptions about the dependence of off-diagonal correlations on $\mathbf{R}$ for each symmetry.  Following \cite{Milstein2003}, we then Fourier transform over the relative coordinate:
\begin{align}
\bar{G}_N \! \left( \mathbf{R}, \mathbf{r} \right) & \rightarrow \tilde{G}_N \! \left( \mathbf{R}, \mathbf{k} \right), \nonumber \\
\bar{G}_A \! \left( \mathbf{R}, \mathbf{r} \right) & \rightarrow \tilde{G}_A \! \left( \mathbf{R}, \mathbf{k} \right),
\label{FTNotn}
\end{align}
which are valid because $\bar{G}_N$ and $\bar{G}_A$ are each symmetric with respect to $\mathbf{r}$.  This transform removes a Dirac delta function appearing in the partial differential equation for the anomalous fluctuations.

\begin{figure}
\includegraphics[width=\columnwidth]{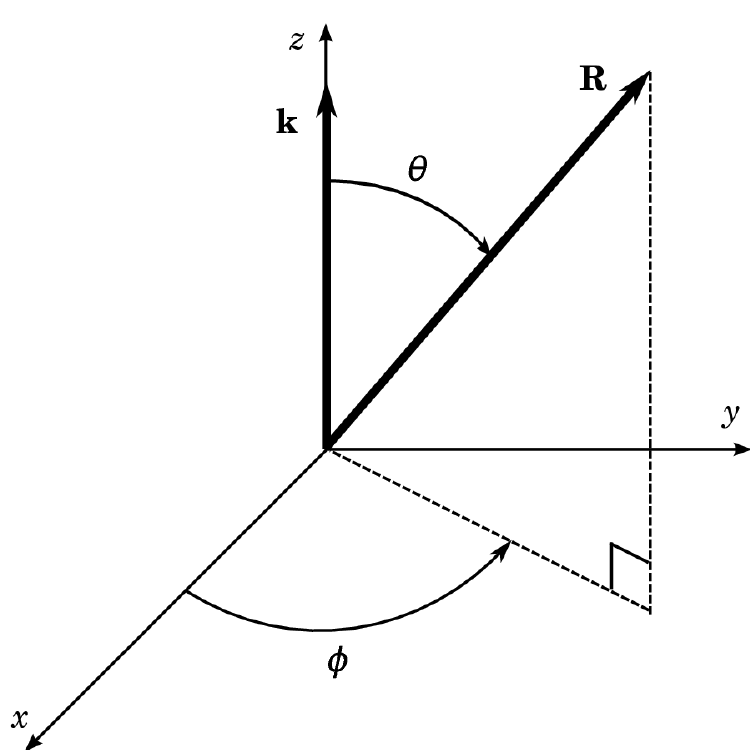}
\caption{Coordinate axes for spherical symmetry.  Since all fields are independent of the orientation of $\mathbf{R}$, we are free to rotate the axes (keeping the origin fixed), and so we align the $z$-axis with $\mathbf{k}$.  Then the dependence of the correlation functions' values on the relative orientations of $\mathbf{k}$ and $\mathbf{R}$ (bold arrows) can be expressed in spherical coordinates.\label{SphericalAxesFig}}
\end{figure}

\begin{figure}
\includegraphics[width=\columnwidth]{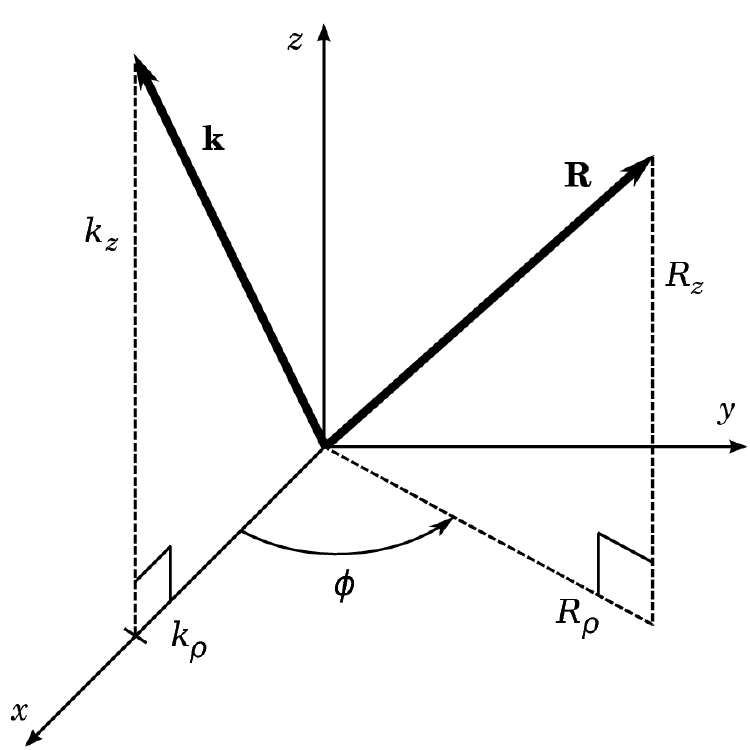}
\caption{Coordinate axes for cylindrical symmetry.  Since all fields are independent of the azimuthal angle of $\mathbf{R}$, we are free to rotate the axes so long as the $z$-axis and origin remain fixed.  We align the $x$-axis with the component of $\mathbf{k}$ lying in the $x$-$y$ plane.  Then the dependence of the correlation functions' values on the relative orientations of $\mathbf{k}$ and $\mathbf{R}$ (bold arrows) can be expressed in cylindrical coordinates.\label{CylindricalAxesFig}}
\end{figure}

Next, we choose coordinate systems appropriate to the geometry.  In the spherically symmetric case, we choose spherical coordinates, where $\mathbf{k}$ is aligned with the $z$-axis, as in Figure \ref{SphericalAxesFig}.
The assumption of spherical symmetry in $\mathbf{R}$ permits this choice, since only the relative orientation of $\mathbf{R}$ and $\mathbf{k}$ will be important.  The angle between $\mathbf{R}$ and $\mathbf{k}$ is then $\theta$, and the azimuthal angle of $\mathbf{R}$ in this coordinate system is $\phi$.  For cylindrical symmetry, we are only permitted to rotate the coordinate axes about the $z$-axis without changing the values of each dependent variable, so we align the $x$-axis with $\mathbf{k_\rho}$, the component of $\mathbf{k}$ lying in the $x\text{-}y$ plane, as in Figure \ref{CylindricalAxesFig}.
The five spatial variables are then $k_z$, $k_\rho$, $R_z$, $R_\rho$, and $\phi$, the axial and radial components of the relative wavenumber and center of mass coordinate, repectively, and the azimuthal angle of $\mathbf{R}$.  The Laplacians and gradients involved are then expressed in spherical or cylindrical coordinates, depending on the geometry; in spherical symmetry, the radial parts of the Laplacians may be simplified, as in the usual treatment of the hydrogen atom, with the substitutions
\begin{align}
\varphi_a \! \left( R \right) & \equiv R \bar{\phi}_a \! \left( R \right), \nonumber \\
\mathcal{G}_N \! \left( R, k, \theta, \phi \right) & \equiv R \tilde{G}_N \! \left( R, k, \theta, \phi \right),
\label{RadDerSphSimp}
\end{align}
and likewise for the molecular field and anomalous fluctuations.  In cylindrical symmetry, we use
\begin{align}
\varphi_a \! \left( R_z, R_\rho \right) & \equiv \sqrt{R_\rho} \bar{\phi}_a \! \left( R_z, R_\rho \right), \nonumber \\
\mathcal{G}_N \! \left( R_z, R_\rho, k_z, k_\rho, \phi \right) & \equiv \sqrt{R_\rho} \tilde{G}_N \! \left( R_z, R_\rho, k_z, k_\rho, \phi \right), \label{RadDerCylSimp}
\end{align}
and likewise for the molecular field and anomalous fluctuations.

Note that, in the spherical case, we have not assumed that the azimuthal angle $\phi$ is unimportant, making our model more general than those used earlier \cite{Milstein2003, Wuster2005}.  Rather than the Legendre polynomial expansions used by Milstein \emph{et al}.\ \cite{Milstein2003} and W\"{u}ster \emph{et al}.\ \cite{Wuster2005}, an appropriate basis for expanding the normal and anomalous fluctuations is then the spherical harmonics $Y_l^q \! \left( \theta, \phi \right)$.  We write
\begin{equation}
\mathcal{G}_{N/A} \! \left( R, k, \theta, \phi \right) = \sum_{l=0}^\infty \sum_{q=-l}^l \mathcal{G}_{N/A}^{l,q} \! \left( R, k \right) Y_l^q \! \left( \theta, \phi \right), \label{GNSphExpand}
\end{equation}
where the N/A subscript means that the equation applies to both $\mathcal{G}_N$ and $\mathcal{G}_A$.  Only one angle is present in the cylindrical case.  Therefore, we use trigonometric functions in $\phi$ to form a complete angular basis, which is common for spectral solutions to PDEs \cite{Fornberg}.  In this respect, sine series are slightly more stringent in their criteria for uniform convergence than cosine series.  Accordingly we expand the normal and anomalous fluctuations as
\begin{multline}
\mathcal{G}_{N/A} \! \left( R_z, R_\rho, k_z, k_\rho, \phi \right) = \\ \sum_{n=0}^\infty \mathcal{G}_{N/A}^n \! \left( R_z, R_\rho, k_z, k_\rho \right) \cos \left( n \phi \right) \label{GNCylExpand}
\end{multline}
where a superscript $n$ indexes a generally complex scalar expansion coefficient and does not denote a power.

The HFB model for spherical symmetry in the center of mass coordinate consists of the denumerably infinite set of equations
\begin{widetext}
\begin{eqnarray}
 \mathrm{i} \hbar \parpar{t} \Rphia{R} &=& \left\lbrace - \frac{\hbar^2}{2m} \parpartwo{R} + \Vabar{R} + U \left[ \frac{\modsq{\Rphia{R}}}{R^2} + 2 \frac{\RGNI{R}}{R} \right] \right\rbrace \Rphia{R} + \left[ U \frac{\RGAI{R}}{R} + g \frac{\Rphim{R}}{R} \right] \Rphiastar{R}, \label{PhiaSphK} \\
\mathrm{i} \hbar \parpar{t} \Rphim{R} &=& \left[ - \frac{\hbar^2}{4m} \parpartwo{R} + \Vmbar{R} + \nu \right] \Rphim{R} + \frac{g}{2} \left[ \frac{\Rphiasq{R}}{R} + \RGAI{R} \right], \label{PhimSphK} \\
 \mathrm{i} \hbar \parpar{t} \RGNlm{l}{q} &=& - \mathrm{i} \frac{\hbar^2 k}{m} \Bigg[ \sqrt{\frac{\left( l-q+1 \right) \left( l+q+1 \right)}{\left( 2l+1 \right) \left( 2l+3 \right)}} \left( \parpar{R} + \frac{l+1}{R} \right) \RGNlm{l+1}{q} \nonumber \\
&& \;\;\;\;\;\;\;\;\;\;\;\;\; + u_{-l+1}^q \sqrt{\frac{\left( l+q \right) \left( l-q \right) }{\left( 2l+1 \right) \left( 2l-1 \right)}} \left( \parpar{R} - \frac{l}{R} \right) \RGNlm{l-1}{q} \Bigg] \nonumber \\
& +& \left( -1 \right)^q \left\lbrace g \frac{\Rphim{R}}{R} + U \left[ \frac{\Rphiasq{R}}{R^2} + \frac{\RGAI{R}}{R} \right] \right\rbrace \RGAlm{* \, l}{-q} \nonumber \\
& - & \left\lbrace g \frac{\Rphimstar{R}}{R} + U \left[ \frac{\Rphiastarsq{R}}{R^2} + \frac{\RGAIstar{R}}{R} \right] \right\rbrace \RGAlm{l}{q}, \label{GNinLM} \\
\mathrm{i} \hbar \parpar{t} \RGAlm{l}{q} &=& \bigg\lbrace - \frac{\hbar^2}{4m} \left[ \parpartwo{R} - \frac{l \left( l+1 \right)}{R^2} - 4k^2 \right] + 2 \Vabar{R} + 4 U \left[ \frac{\modsq{\Rphia{R}}}{R^2} + \frac{\RGNI{R}}{R} \right] \bigg\rbrace \RGAlm{l}{q} \nonumber \\
& +& \left\lbrace g \frac{\Rphim{R}}{R} + U \left[ \frac{\Rphiasq{R}}{R^2} + \frac{\RGAI{R}}{{R}} \right] \right\rbrace \left[ \RGNlm{l}{q} + \left( -1 \right)^q \RGNlm{* \, l}{-q} + \sqrt{4 \pi} \delta_{0,l} \, \delta_{0,q} R \right], \label{GAinLM}
\end{eqnarray}

\end{widetext}
where the diagonal elements $\mathcal{G}_{N/A}^I$ of the normal and anomalous fluctuations, respectively, are computed from
\begin{equation}
\mathcal{G}_{N/A}^I \! \left( R \right) = \frac{1}{4 \pi^{5/2}} \int_0^\infty \mathrm{d}k \, k^2 \mathcal{G}_{N/A}^{0,0} \! \left( R, k \right).
\label{GNISphDefn}
\end{equation}
The discrete step function $u_a^b$ is 1 if $a \geq b$ and 0 otherwise.  In the cylindrical case, the denumerably infinite set of equations is
\begin{widetext}
\begin{eqnarray}
 \mathrm{i} \hbar \parpar{t} \Rphia{\Rcyls} &= & \Bigg\lbrace - \frac{\hbar^2}{2m} \left( \frac{1}{4 R_{\rho}^2} + \parpartwo{R_{\rho}} + \parpartwo{R_z} \right) + \Vabar{\Rcyls}
 + U \left[ \frac{\modsq{\Rphia{\Rcyls}}}{R_{\rho}} + 2 \frac{\RGNI{\Rcyls}}{\Rhalf} \right] \Bigg\rbrace \nonumber \\
 &\times& \Rphia{\Rcyls}
 + \left[ U \frac{\RGAI{\Rcyls}}{\Rhalf} + g \frac{\Rphim{\Rcyls}}{\Rhalf} \right] \Rphiastar{\Rcyls}, \label{PhiaCylR}\\
 \mathrm{i} \hbar \parpar{t} \Rphim{\Rcyls} &= & \left[ -\frac{\hbar^2}{4m} \left( \frac{1}{4 R_{\rho}^2} + \parpartwo{R_{\rho}} + \parpartwo{R_z} \right) + \Vmbar{R_{\rho},R_z} + \nu \right] \Rphim{\Rcyls} \nonumber \\
 &+& \frac{g}{2} \left[ \frac{\Rphiasq{\Rcyls}}{\Rhalf} + \RGAI{\Rcyls} \right], \label{PhimCylR}\\
\hbar \parpar{t} \RGNn{n} &= & - \frac{\hbar^2}{2m} \bigg[ k_\rho \left( \parpar{R_\rho} + \frac{2n+1}{2 R_\rho} \right) \RGNn{n+1} \nonumber \\
&& \;\;\;\;\;\;\;\;\;\;\;\; + k_\rho \left( \parpar{R_\rho} - \frac{2n-1}{2 R_\rho} \right) u_1^n \, \RGNn{n-1} \nonumber \\
&& \;\;\;\;\;\;\;\;\;\;\;\; + k_\rho \left( \parpar{R_\rho} - \frac{1}{2 R_\rho} \right) \delta_{n,1} \, \RGNn{0} + 2 k_z \parpar{R_z} \RGNn{n} \bigg] \nonumber \\
& +& 2 \, \text{Im} \! \left( \left\lbrace g \frac{\Rphim{\Rcyls}}{R_\rho^{1/2}} + U \left[ \frac{\Rphiasq{\Rcyls}}{R_\rho} + \frac{\RGAI{\Rcyls}}{R_\rho^{1/2}} \right] \right\rbrace
\RGAn{n*} \right), \\
\mathrm{i} \hbar \parpar{t} \RGAn{n} &= & \bigg\lbrace - \frac{\hbar^2}{4m} \left[ \frac{1-4n^2}{4 R_\rho^2} + \parpartwo{R_\rho} + \parpartwo{R_z} - 4 \left( k_\rho^2 + k_z^2 \right) \right] \nonumber \\
&& \;\;\;\; + 2 \Vabar{\Rcyls} + 4 U \left[ \frac{\modsq{\Rphia{\Rcyls}}}{R_\rho} + \frac{\RGNI{\Rcyls}}{R_\rho^{1/2}} \right] \bigg\rbrace \RGAn{n} \nonumber \\
& +& \left\lbrace g \frac{\Rphim{\Rcyls}}{R_\rho^{1/2}} + U \left[ \frac{\Rphiasq{\Rcyls}}{R_\rho} + \frac{\RGAI{\Rcyls}}{R_\rho^{1/2}} \right] \right\rbrace \nonumber \\
&& \times \left[ \RGNn{n} + \RGNn{n*} + R_\rho^{1/2} \delta_{n,0} \right], \label{GAinn}
\end{eqnarray}

\end{widetext}
with
\begin{multline}
\RGNAI{\Rcyls} = \frac{1}{\left( 2 \pi \right)^2} \int_{-\infty}^{\infty} \wrt{k_z} \nonumber \\
\int_0^{\infty} k_\rho \, \wrt{k_\rho} \RGNAn{0}.
\label{GNIinn}
\end{multline}
Note that the normal fluctuations will be real in either symmetry if they are initially real.

\begin{figure}
\includegraphics[width=\columnwidth]{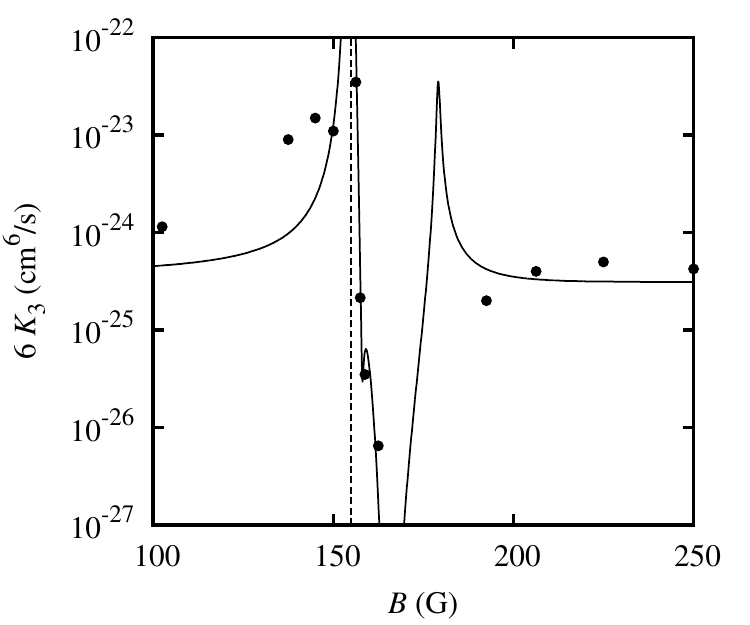}
\caption{A comparison of the measured \cite{Roberts2000} (circles) and analytical (solid line) three-body loss rate of Eq.\ (\ref{K3Defn}) as a function of magnetic field.  Eq.\ (\ref{K3Defn}) has been multiplied by a factor of $6$ here, because the measurements were done in a thermal gas.  The dashed vertical line marks $B = 154.9$ Gauss, where the effective scattering length diverges.\label{K3FitFig}}
\end{figure}

Where desired, we can incorporate a three-body loss coefficient into our spherical model for the atoms inside the atomic condensate, by fitting the analytical model used by Braaten and Hammer \cite{BraatenHammer2006} and D'Incao \emph{et al}.\ \cite{D'IncaoEtAl2009} to the empirical data of Roberts \emph{et al}.\ \cite{Roberts2000}.  This model assumes that universal Efimov physics, in combination with the weakly-bound Feshbach state and deeply bound dimer states, give rise to the three-body recombination.  The result is the addition of the term
\begin{equation}
-\frac{1}{2} \mathrm{i} \hbar K_3 \! \left( a_\text{eff} \right) \frac{\left| \Rphia{R} \right|^4}{R^4} \Rphia{R}
\label{SphK3Term}
\end{equation}
to the right hand side of Eq.\ (\ref{PhiaSphK}), where
\begin{widetext}
\begin{equation}
K_3 \left( a_\text{eff} \right) = \frac{\hbar}{2m} a_\text{eff}^4
\begin{cases}
67.1 \cdot \mathrm{e}^{-2 \eta} \left[ \cos^2 \left( s_0 \ln a_\text{eff}/a_\text{pos} \right) + \sinh^2 \eta \right] + 16.8 \left( 1 - \mathrm{e}^{-4 \eta} \right) & \text{ if } a_\text{eff} > 0 \\
4590 \left( \sinh 2\eta \right)/\left[ \sin^2 \left( s_0 \ln a_\text{eff}/a_\text{neg} \right) + \sinh^2 \eta \right] & \text{ if } a_\text{eff} < 0
\end{cases}
\Bigg\rbrace,
\label{K3Defn}
\end{equation}
\end{widetext}
with $s_0 = 1.00624$, $\eta = 0.0165971$, $a_\text{pos} = 236.743 a_0$, and $a_\text{neg} = - a_\text{pos}/0.96$.  Figure \ref{K3FitFig} compares this loss rate with measured values \cite{Roberts2000}.  The leading scaling goes as $a_\text{eff}^4$, which has been shown to be effective elsewhere \cite{SmirneEtAl2007}.  In practice, we impose an empirically based floor of $10^{-28}$ cm$^6$/s on $K_3 \! \left( a_\text{eff} \right)$.

\section{\label{ResultsSec}Results of HFB Simulations}

Our simulations are performed in spherical symmetry using the method of lines \cite{Schiesser}.  We do not perform more realistic axisymmetric simulations, since they would require a larger amount of time.  The infinite sums over $l$ in Eqs.\ (\ref{GNinLM}) and (\ref{GAinLM}) are truncated at $l = 1$, which is the smallest value that provides satisfactorily converged results.  Truncating at $l = 3$ does not significantly affect the results (for example, atom number during a collapse simulation changes by less than $0.002$ percent and by less than $0.003$ percent during the longest, most repulsive pulse simulation), though computation time is significantly larger.  The expansion coefficients for any nonzero $q$ are coupled only to those for $-q$, and only $q=0$ is necessary for calculation of the diagonal elements of the fluctuations.  Therefore, we only propagate $q=0$ in our simulation, and forego whatever information about fluctuations is present in the $q \ne 0$ coefficients.  A fifth order Runge-Kutta method provides time propagation, and we handle the spatial derivatives by spectral collocation in a sinusoidal basis.  We neglect three-body effects unless noted otherwise: as we will show, three-body effects do not significantly alter our results.

We create initial states (non-interacting in attractive simulations and with $a_\text{eff} = 7 a_0$ in repulsive simulations) using imaginary time propagation with the Gross-Pitaevskii equation in the trap used in Ref.\ \cite{Donley2001}, using $\omega = 2 \pi \times 12.77$ radians per second, which is the geometric mean of the experimental trapping frequencies.  Imaginary time propagation is necessary because the simulation's implementation imposes infinite potential barriers at the end of the grid, so that a closed-form solution for the Gross-Pitaevskii equation is not available.  We assume the molecular field's initial state is
\begin{equation}
\phi_m = - \frac{g}{2 \nu} \phi_a^2,
\label{IniPhim}
\end{equation}
which is the exact initial state for a uniform gas without fluctuations.  Noting the importance of the molecular binding energy \cite{Kokkelmans2002a}, we always choose $g_c = 1.816$, except where noted.

\subsection{\label{AttrSubSec}Attractive Interactions}

The collapse simulation begins with the effective scattering length set to $-12 a_0$, where it remains for the duration of the simulation.  All simulation parameters are taken as the physical parameters of the experiment of Ref.\ \cite{Donley2001}.

\begin{figure}
\includegraphics[width=\columnwidth]{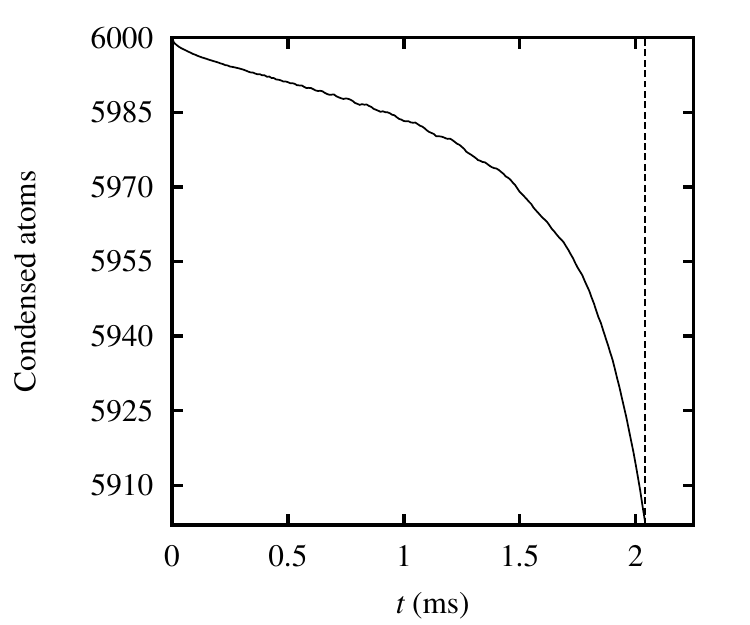}
\caption{Number of condensed atoms in the collapse simulation.  The dashed vertical line marks $t = 2.04$ ms, the last sample point for which we simulation is stable.\label{ColNaFig}}
\end{figure}

\begin{figure}
\includegraphics[width=\columnwidth]{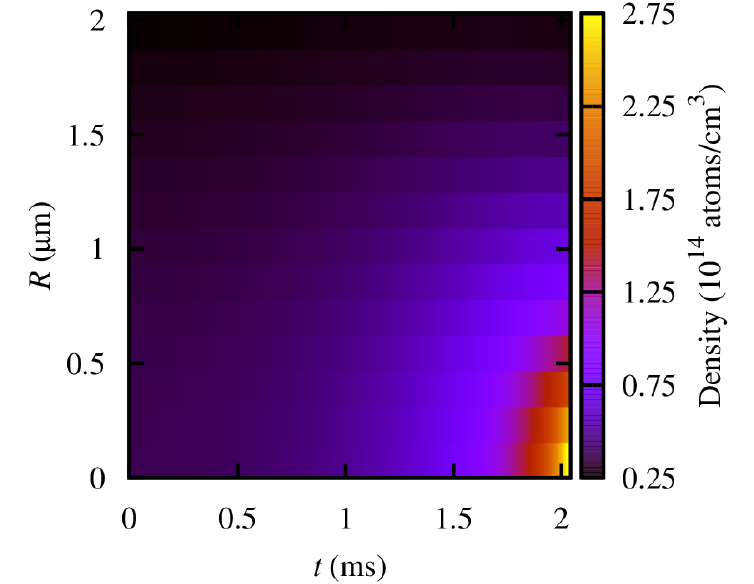}
\caption{(Color online) Condensed atom density in the collapse simulation.  In the simulation, $R$ extends to $10$ $\mu$m.\label{ColADensFig}}
\end{figure}

\begin{figure}
\includegraphics[width=\columnwidth]{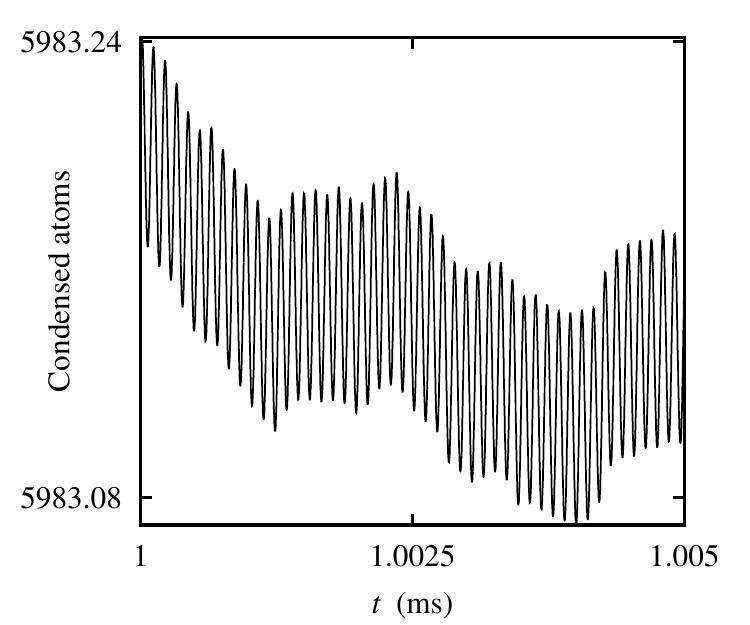}
\caption{Number of condensed atoms during a brief period in the middle of the collapse simulation.  The plotted points are $0.5$ nanoseconds apart, so the visible oscillations of period $1.1$ microseconds are not aliased.\label{ColNaZoomFig}}
\end{figure}

\begin{figure}
\includegraphics[width=\columnwidth]{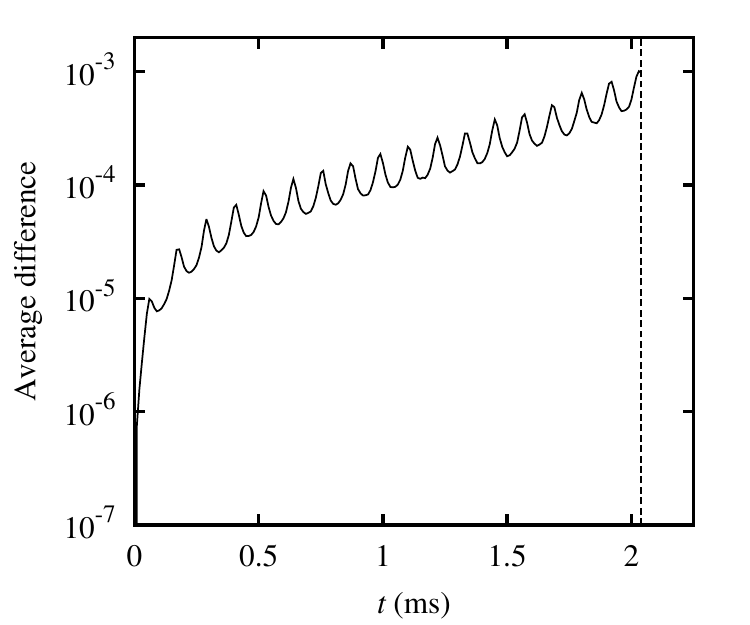}
\caption{Difference over the average of the number of condensed atoms between collapse simulations with and without three-body effects.  The dashed vertical line marks $t = 2.04$ ms, the last sample point for which the simulation is stable.  At no time does the average difference exceed $0.2$ percent, though it is consistently increasing.\label{NaDiffColFig}}
\end{figure}

Figure \ref{ColNaFig} plots the number of atoms in the condensate as a function
of time, and Figure \ref{ColADensFig} shows the condensate density.  The erratic
oscillations around the general downward trend visible in Figure \ref{ColNaFig}
are more clearly shown in Figure \ref{ColNaZoomFig}.  We interpret $2$
milliseconds as our predicted collapse time, because the condensate number
begins to drop substantially and the density grows increasingly concentrated
near the origin at about this time.  This value can be directly compared to the
nearest data point in Donley \emph{et al}.'s Figure 2 \cite{Donley2001}, which
has a collapse time of about $2 \pm 1$ milliseconds.  Beyond $2.04$
milliseconds, negative noncondensed densities consistently start to appear in
our simulation, and the error in total number diverges.  Such instability may
occur because the spatial grid is uniform and non-adaptive, while most of the
dynamics occur on the innermost grid points once the collapse has begun.  Thus,
an adaptive spatial grid would be required to be sure of a $t_\text{collapse}$
value more precise than $2$ milliseconds.  Including three-body effects changes
the results by less that $0.2$ percent, as Figure \ref{NaDiffColFig} shows.  The
$K_3$ factor in the collapses is at the floor value of $10^{-28}$ cm$^6$/s,
which is twice the maximum value that Altin \emph{et al}.\ \cite{AltinEtAl2011}
find allowable.  However, since the loss term scales as the square of the
density, and Altin \emph{et al}.'s condensates are several times denser than
those of the JILA experiments, it seems likely that $K_3$ would indeed have to be much larger to have a noteworthy effect in our simulations.

\begin{figure}
\includegraphics[width=\columnwidth]{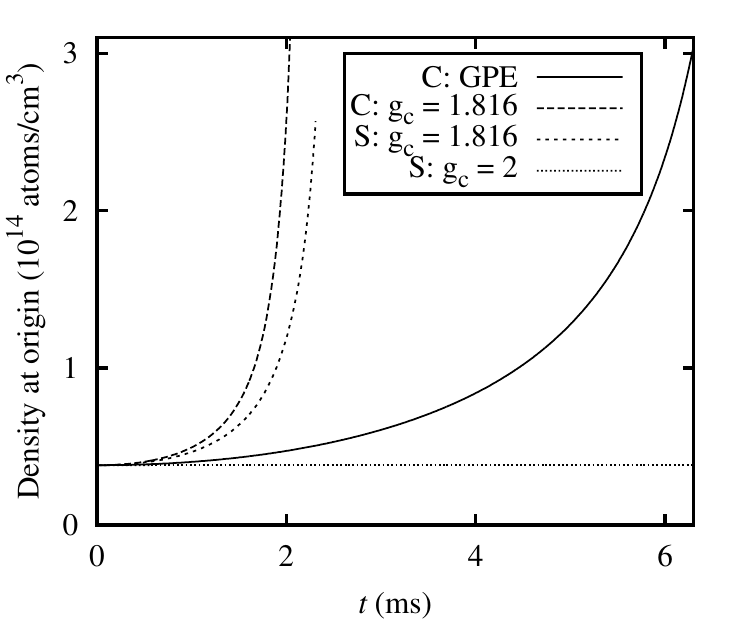}
\caption{Density of atoms in the atomic condensate as predicted by the Gross-Pitaevskii equation (solid line) during a simulated collapse and by various configurations of our model.  For readability, the abscissa here is truncated before the GPE simulation ends; its density at the origin climbs until instability terminated the simulation at $7.35$ milliseconds.  Our model with $g_c = 1.816$ (long dashed line) collapses much earlier than the GPE for a given magnetic field, but also collapses for $B = B_\text{res} + \Delta B$ (short dashed line).  For $g_c = 2$, our model predicts a rise in density very similar to the GPE (a plot would overlap the GPE result shown here), and no effective change in density when $B = B_\text{res} + \Delta B$ (dotted line).  The ``C" and ``S" labels in the legend are reminders as to which series are for nominally collapsing and stationary values of $a_\text{eff}$, respectively.\label{GPECompareCollapse}}
\end{figure}

These results are compared to mean field theoretic results in Figure \ref{GPECompareCollapse}, which plots atomic condensate density at the origin as calculated from simulations using our model and the GPE.  A comparison of numbers of atoms would be pointless, since the GPE only allows particles to be in the atomic condensate.  Characterizing collapse by density at the origin, the GPE predicts a collapse time more than $5$ milliseconds later than our model predicts and the experiments measure.  Figure \ref{GPECompareCollapse} also shows the results obtained with our model when $B = B_\text{res} + \Delta B$.  First, we use the nominal value of the coupling correction factor $g_c = 2$.  According to Equation (\ref{aEffFromNu0}), the scattering length $a_\text{eff} = 0$ for this field.  Then, both the GPE simulation and our model simulation result in a stationary solution, as would be expected for the noninteracting case.  When we set the scattering length to $-12 a_0$, our model gives results nearly identical to the GPE on atom number and condensate density at the origin (see Figure \ref{GPECompareCollapse}), and a collapse occurs in both simulations.  These results are consistent with the findings in Ref.\ \cite{Wuster2005}.  However, when we tune the coupling correction factor to the value $g_c = 1.816$ to match the correct molecular binding energy, the simulation results of our model differ significantly from the nominal GPE results.  Simulation results are again shown in Figure \ref{GPECompareCollapse} for $B = B_\text{res} + \Delta B$, but now we observe a collapse.  This can be understood again from Equation (\ref{aEffFromNu0}), since at this field for $g_c = 1.816$ we obtain the value $a_\text{eff} = -41 a_0$.

As we discussed before, it is not possible in our model to have the binding energy and the mean field energy both correctly described, which leads to a trade-off in what is the dominant effect causing the collapse.  For our treatment of the collapse problem, we argue that the collapse near a Feshbach resonance is most sensitive to the binding energy; however, since we have to set accordingly the value $g_c = 1.816$, this unfortunately leads to stability problems in the zero-crossing region of the scattering length.

The molecular field is very sparsely populated, never totalling more that $10$ percent of a molecule, even though pair formation is the primary mechanism in the model \cite{Kokkelmans2002} of interatomic interactions that we used.  However, this does not mean the molecular field could simply be neglected: in a two-channel model of a Feshbach resonance the molecular field often has very small occupation but plays a key role in both static and dynamic descriptions of the physical molecule, a superposition between the atomic (open) and molecular (closed) channels.  The oscillations between atoms and molecules, especially visible in Figure \ref{ColNaZoomFig}, have a period of $1.1$ microseconds, which is approximately equal to $t_U = m^3 U^2/\hbar^5$, a natural time scale arising from dimensional analysis of the parameters of the model.  Other analytically deduced time scales are $t_g = \hbar^7/m^3 g^4 \approx 0.57$ nanoseconds, and $t_\nu = \hbar/\nu \approx 22$ nanoseconds.

\begin{figure}
\includegraphics[width=\columnwidth]{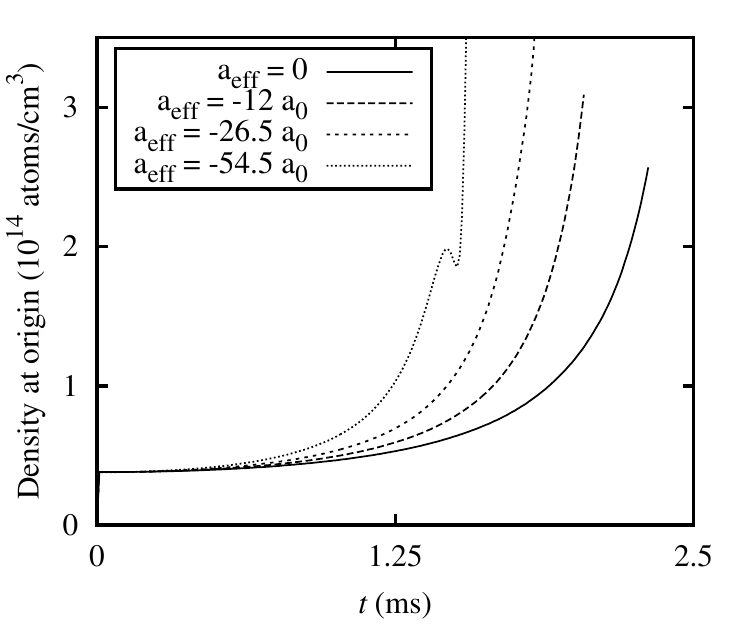}
\caption{Atomic condensate density at the origin for simulations with effective scattering lengths of $0$ (solid line), $-12 a_0$ (long dashes), $-26.5 a_0$ (short dashes), and $-54.5 a_0$ (dotted line).  These simulations become unstable after $2.31$, $2.04$, $1.88$, and $1.6$ milliseconds, respectively.  All simulations use $g_c = 1.816$.  \label{OtherAEffFig}}
\end{figure}

While most of our collapse results are for $a_\text{eff} = -12 a_0$, Figure \ref{OtherAEffFig} shows atomic condensate density at the origin for other values of $a_\text{eff}$.  We find that for $a_\text{eff} = -26.5 a_0$ and $a_\text{eff} = -54.5 a_0$, respectively, the simulations become unstable after $1.88$ milliseconds and $1.6$ milliseconds, respectively.  These times are within the error bounds for the corresponding experimental collapse times reported in Figure 2 of Ref.\ \cite{Donley2001}, and correctly capture the general trend of collapse time decreasing nontrivially as effective scattering length becomes more negative.

\subsection{\label{RepuSubSec}Repulsive Interactions}

\begin{figure}
\includegraphics[width=\columnwidth]{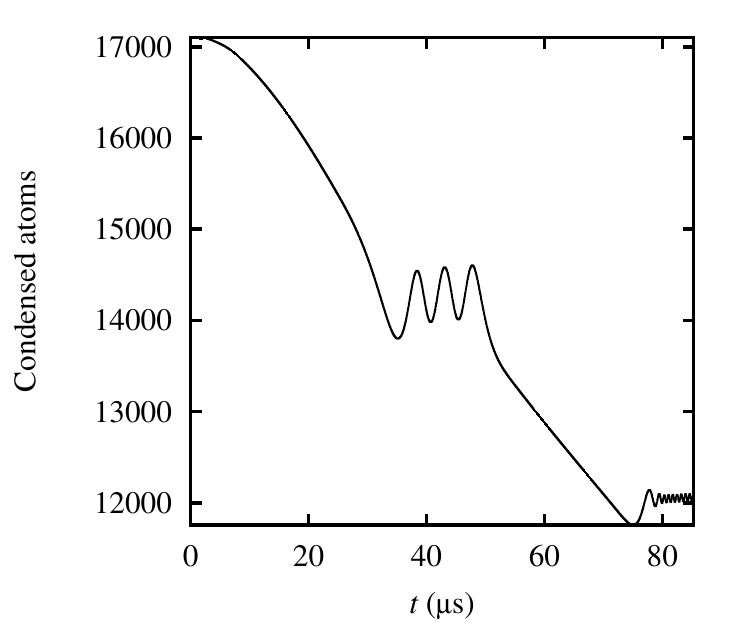}
\caption{Number of condensed atoms in the two-pulse simulation.  The oscillations in the middle of the simulation occur during the free precession time after the first pulse and before the second, when the magnetic field is held constant.\label{p2NaFig}}
\end{figure}

It is interesting to compare our nonhomogeneous model with the double pulse experiment of Donley {\it et al.}~\cite{Donley2002}, which has been described succesfully earlier with a similar but homogeneous model~\cite{Kokkelmans2002a} by applying the local density approximation. In this experiment, a condensate is subjected to two magnetic field pulses with interatomic interactions that become very repulsive.  Although we do not expect as large inhomogeneity effects as in the attractive situation where a collapse occurs, it is useful to compare the results between the models as it indicates the validity of the local density approximation. The results are indeed different.  With an initial state having $16,600$ atoms, all in the condensate, and a free evolution time of 10 microseconds (the left-most set of data points in Figure 5 of Ref.\ \cite{Kokkelmans2002a}), we find about $12,000$ condensed atoms and slightly over $4,500$ noncondensed atoms at the end of the simulation.  Kokkelmans and Holland~\cite{Kokkelmans2002a} find about $9,500$ and $6,300$, respectively, while the experiment \cite{Donley2002} ends with just over $10,000$ condensed atoms and slightly under $5,000$ noncondensed atoms.  Figure \ref{p2NaFig} shows the number of condensed atoms over the course of our simulation.  It should be noted that a homogeneous version of our code exactly reproduced the results of the code used in Ref.\ \cite{Kokkelmans2002a}.

\begin{figure}
\includegraphics[width=\columnwidth]{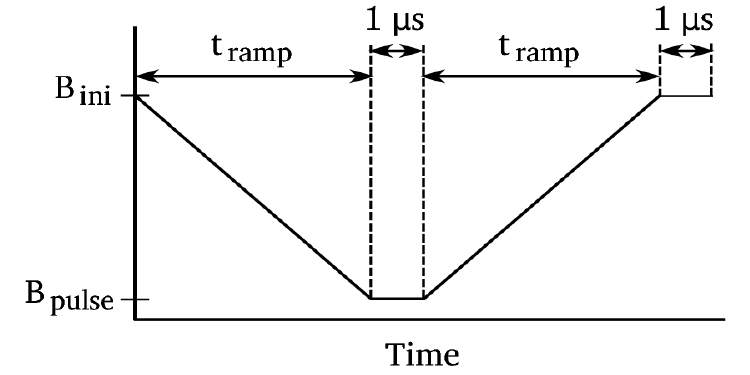}
\caption{Time dependence of the magnetic field for simulations with a single pulse that induces repulsive interatomic interactions.  The lowest value of the magnetic field is $B_\text{pulse}$, and the time to descend from the initial value $B_\text{ini} \approx 165.68$ Gauss, where $a_\text{eff} = 7 a_0$, to $B_\text{pulse}$ is called $t_\text{ramp}$.  Our simulations had the magnetic field near resonance for $1$ microsecond, though experiments were conducted \cite{Claussen2002} with many other values.\label{OnePulseFig}}
\end{figure}

\begin{figure}
\includegraphics[width=\columnwidth]{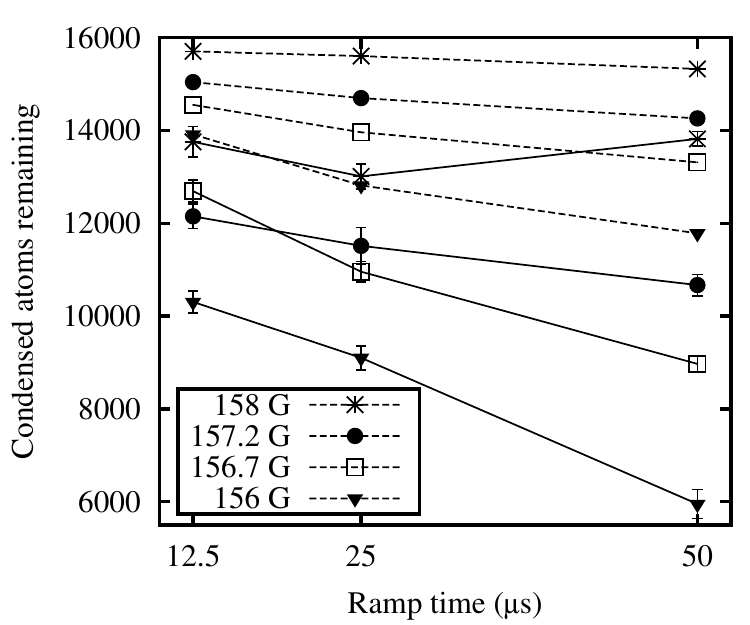}
\caption{Numbers of condensed atoms remaining at the ends of simulations (glyphs connected by dashed lines) and experiments \cite{Claussen2002, ClaussenPrivCom} (glyphs connected by solid lines) as a function of $t_\text{ramp}$.  Curves are provided as a guide to the eye; actual data is represented by glyphs.\label{p1ExptCompFig}}
\end{figure}

\begin{figure}
\includegraphics[width=\columnwidth]{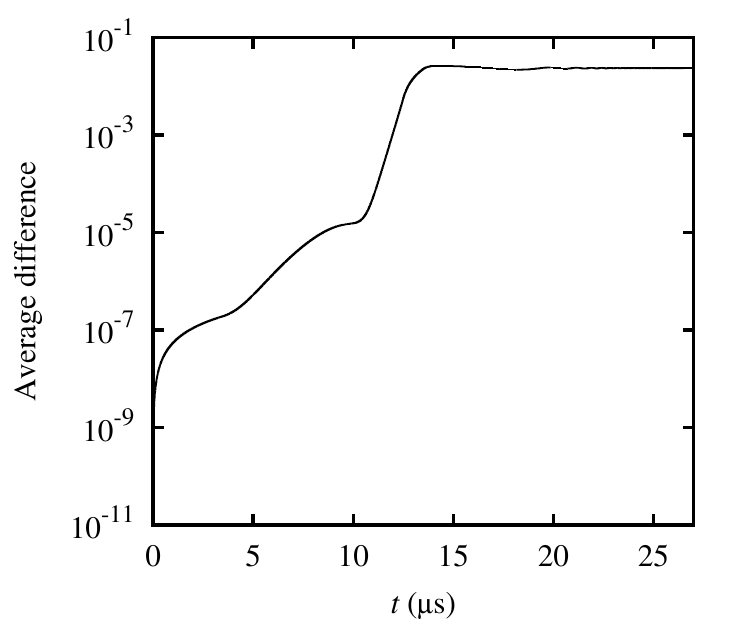}
\caption{Difference over the average of the number of condensed atoms between simulations with and without three-body effects.  The scenario was a single pulse to $B = 156$ Gauss.  At no time does the average difference exceed $2.7$ percent, though it increases most dramatically when the magnetic field is near resonance.\label{NaDiffPulseFig}}
\end{figure}

In comparison with a second type of experiment, we conducted simulations using
the same parameters as those used to create the left-most three sets of data
points in Figure 4 of Ref.\ \cite{Claussen2002}, in which a condensate is
subjected to a single brief magnetic field pulse of strength $B_\text{pulse}$.
Both the pulse strength and the time $t_\text{ramp}$ to reach that value of
magnetic field are varied between experiments.  Figure \ref{OnePulseFig} shows
the time dependence of the magnetic field used in the simulations.  The results
are plotted alongside Claussen \emph{et al}.'s \cite{ClaussenPrivCom} in Figure
\ref{p1ExptCompFig}, which includes experimental error bars.  Error bars on the
simulated data points, which would reflect the differences between different
grid resolutions, are too small to see in the figure.  Our simulations
consistently end with higher numbers of condensed atoms than the experiments
did, and the condensed atom number decreased monotonically with longer ramp
times.  However, the experimental results are not easy to interprete. For
instance, there is a local minimum in the $158$ Gauss series which is not
observed in the simulations.  Moreover, the experimental data have a larger
number of atoms remaining for the $156.7$ Gauss pulse than the $157.2$ Gauss
pulse for the shortest ramp time; our simulations do not reproduce this effect,
either.  Including three-body effects does not significantly change our results,
as Figure \ref{NaDiffPulseFig} shows.  Again, the $K_3$ term in these
simulations is larger than what Altin \emph{et al}.\ predict, but note that
these simulations are for relatively short times, and for condensates that are
over an order of magnitude sparser than those of Altin \emph{et al}.\ and
expanding further.

\begin{figure}
\includegraphics[width=\columnwidth]{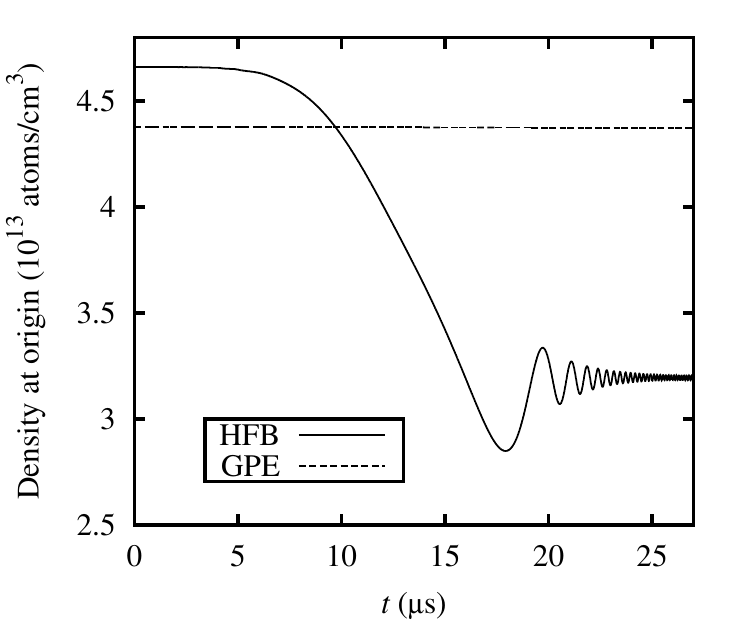}
\caption{Density of atoms in the atomic condensate as predicted by our model (solid line) and the Gross-Pitaevskii equation (dashed line) during a simulation of a pulse to $B_\text{pulse} = 156$ Gauss.\label{GPEComparePulse}}
\end{figure}

Our model's results for atomic condensate density at the origin are compared to those of the GPE in Figure \ref{GPEComparePulse} for the $B_\text{pulse} = 156$ Gauss case.  Not only does the GPE predict an invariant number of condensed atoms, but it also predicts a nearly invariant density.  This comparison is representative of all of our simulations with repulsive interactions.

\begin{figure}
\includegraphics[width=\columnwidth]{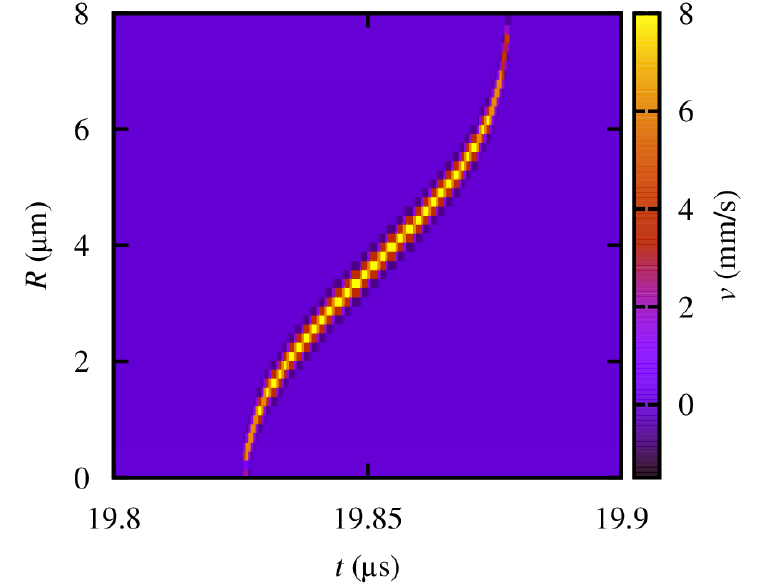}
\caption{(Color online) Velocities of molecules during a portion of a simulation involving a pulse to $156$ G.\label{p1mvStreak}}
\end{figure}

Figure \ref{p1mvStreak} shows the molecular velocities during a small portion of the $B_\text{pulse}=156$ Gauss simulation, calculated as the gradient of the phase.  Such instances of very high velocities appear periodically, when the molecular field declines.  We infer that the molecular binding energies are transformed to kinetic energy as the molecules dissociate.

\begin{figure}
\includegraphics[width=\columnwidth]{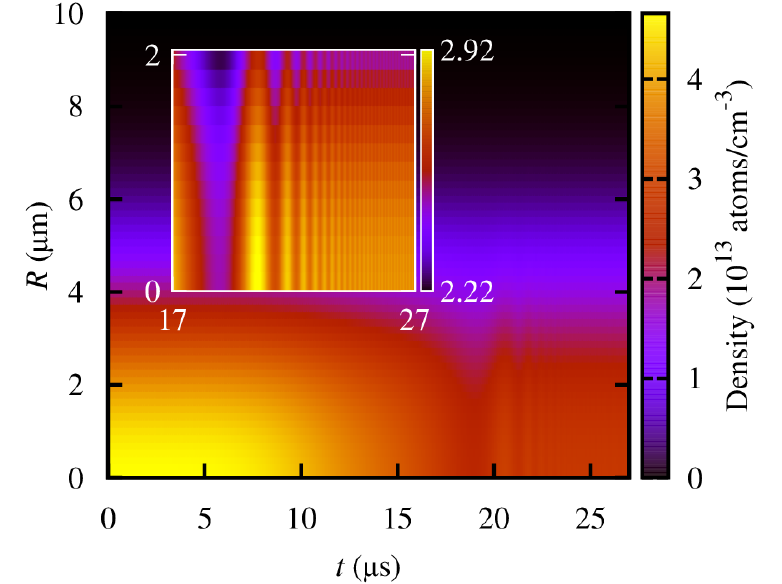}
\caption{(Color online) Condensed atom density during a pulse to $B_\text{res}$.  The inset magnifies the region where the oscillations in density are most pronounced.\label{p1aDensFig}}
\end{figure}

\begin{figure}
\includegraphics[width=\columnwidth]{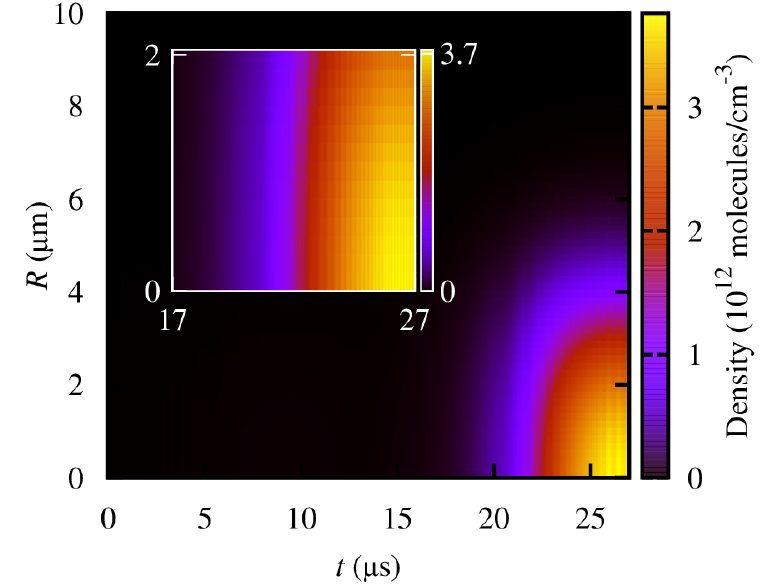}
\caption{(Color online) Molecule density during a pulse to $B_\text{res}$.  The inset magnifies the same region as the inset in Figure \ref{p1aDensFig}.\label{p1mDensFig}}
\end{figure}

\begin{figure}
\includegraphics[width=\columnwidth]{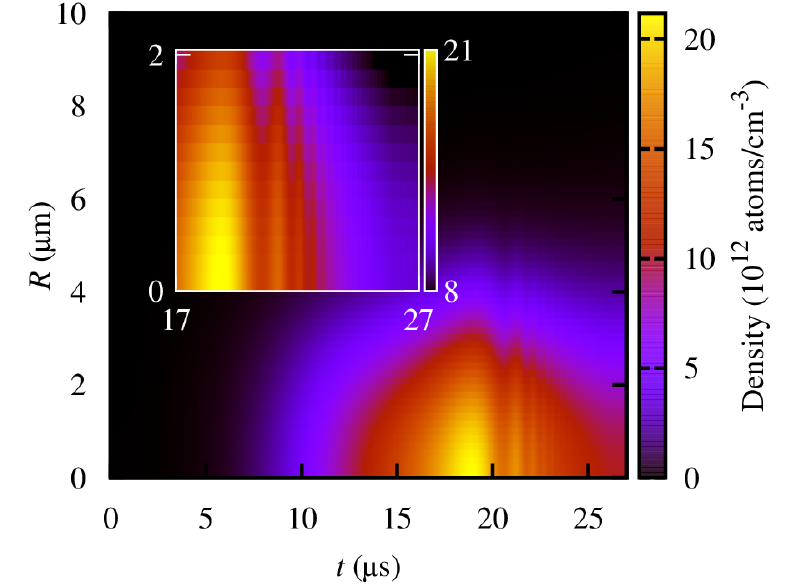}
\caption{(Color online) Noncondensed atom density during a pulse to $B_\text{res}$.  The inset magnifies the same region as the inset in Figure \ref{p1aDensFig}.\label{p1nDensFig}}
\end{figure}

We also performed simulations with $B_\text{pulse} = B_\text{res}$.  Note that the effective scattering length diverges for this value of the magnetic field, and the Gross-Pitaevskii equation becomes undefined.  The results follow the trend established in Figure \ref{p1ExptCompFig}, with fewer than $10,000$ condensed atoms remaining when $t_\text{ramp} = 50$ microseconds.  Figures \ref{p1aDensFig}, \ref{p1mDensFig}, and \ref{p1nDensFig} show the number densities of condensed atoms, molecules, and noncondensed atoms, respectively, for the simulation with $B_\text{pulse} = B_\text{res}$ and $t_\text{ramp} = 12.5$ microseconds.

\section{\label{ConcSec}Conclusion}

We have derived a two-channel model of BECs in which the constituent atoms
interact via a Feshbach resonance.  The model includes first order fluctuations
around the atomic mean field and is computationally feasible in spherical or
cylindrical symmetry.  We compare our model to the results of the collapse
experiments at JILA, which so far always have been difficult to explain by
theory.

The model, which accounts for interactions using a mean field of bound pairs, approximates the experimentally measured time to collapse in a simulation of a BEC with attractive interactions, but only when the renormalization incorporates the correct molecular binding energy at the cost of incorporating an incorrect mean field energy.  It is important to note that our model cannot simulate the whole regime between an unstably attractive gas and a noninteracting gas with a single value of the parameter $g_c$.

The inclusion of three-body effects does not significantly affect the
condensate's dynamics leading up to collapse, in contrast to recent experimental
findings at the Australian National University \cite{AltinEtAl2011}, which operated at much higher densities.  Three-body effects may increase the magnitude of atom loss, especially near and after $t_\text{collapse}$, when the condensate's density is very high.  Near and after the collapse time, the density is strongly localized at the origin, and our simulations cease to be numerically tractable due to our choice of a regular grid.  A finite element approach or other alternative grid may allow post-collapse simulations in the future.  Such an approach could be particularly effective if utilized in the center of mass coordinate.

The model does not quantitatively reproduce the results of experiments with a single brief period of repulsive interactions.  In our simulations of these experiments, a significant number of molecules are formed, concurring with expectations \cite{Claussen2002, Donley2002, Kokkelmans2002a} that a nontrivial molecular condensate is formed and coexists with the atomic condensate under these circumstances.  We found that the use of three-body recombination did not improve matters.  However, the large number of molecules strains the assumption in our model that fluctuations around the molecular field are negligible.  Incorporating normal and anomalous molecular fluctuations would add considerably to the computational requirements.  Interactions of molecules with atoms and with each other, such as collisions, may temper the high velocities observed in our simulations, possibly resulting in a larger noncondensed component and smaller condensed fields.

\section*{Acknowledgements}
We thank Neil Claussen, Murray Holland, and Carl Weiman for useful discussions.  S.J.J.M.F.K. acknowledges support from the Netherlands Organization for Scientific Research (NWO).  This material is based in part upon work supported by the National Science Foundation under grant numbers PHY-0547845, PHY-0903457, and PHY-1067973; LDC also acknowledges support from AFOSR and the Alexander von Humboldt foundation.  We also acknowledge the Golden Energy Computing Organization at the Colorado School of Mines for the use of resources acquired with financial assistance from the National Science Foundation and the National Renewable Energy Laboratory.

\bibliography{SKC2011}

\end{document}